\documentclass[a4paper,10pt,twoside]{cpc-hepnp}

\usepackage{multicol}
\usepackage{graphicx}
\usepackage{booktabs}
\usepackage{amssymb,bm,mathrsfs,bbm,amscd}
\usepackage[tbtags]{amsmath}
\usepackage{lastpage}
\usepackage{leftidx}
\usepackage{CJK}

\begin{document}
\begin{CJK*}{GBK}{song}

\fancyhead[c]{\small Submitted to 'Chinese Physics C'}


\title{${^{13}}C\left(\alpha,n\right){^{16}}O$ background in a liquid scintillator based neutrino experiment }


\author{%
      Jie Zhao$^{1,2}$
\quad Ze-Yuan Yu$^{2;1)}$\email{yuzy\oa ihep.ac.cn}
\quad Jiang-Lai Liu$^{3}$
\quad Xiao-Bo Li$^{2}$\\
\quad Fei-Hong Zhang$^{1,2}$
\quad Dongmei Xia$^{1,2}$
}

\maketitle

\address{%
$^1$ (University of Chinese Academy of Sciences, Beijing 100049, China)\\
$^2$ (Institute of High Energy Physics, Chinese Academy of Sciences, Beijing 100049, China)\\
$^3$ (Shanghai Jiaotong University, Shanghai 200240, China)\\
}

\begin{abstract}
$\alpha$ from natural radioactivity may interact with a nucleus and emit a neutron. The reaction introduces background to the liquid scintillator (LS) based neutrino experiments. In the LS detector, $\alpha$ comes from ${^{238}}U$, ${^{232}}Th$,  and ${^{210}}Po$ decay chains. For Gadolinium-doped LS (Gd-LS) detector, $\alpha$ also comes from ${^{227}}Ac$. The nucleus ${^{13}}C$ is a natural component of Carbon which is rich in the LS. The background rate and spectrum should be subtracted carefully from the neutrino candidates. This paper describes the calculation of neutron yield and spectrum with uncertainty estimated.The results are relevant for many existing neutrino experiments and future LS or Gd-LS based experiments.
\end{abstract}




\begin{multicols}{2}

\section{Introduction}

Neutrino oscillation is one of the keys to new physics beyond the Standard Model. In recent years, the role of reactor neutrino experiments in the precise determination of the oscillation parameters has become increasingly important, as signified in the results of KamLAND~\cite{kamland}, DayaBay~\cite{dybprl}, RENO~\cite{reno} and Double Chooz~\cite{doublechooz}. Generally, these experiments utilize Gd-LS or LS as target, detecting reactor electron antineutrinos ($\bar \nu_e$) via inverse $\beta$ decay: $\bar \nu_e+p\rightarrow e^++n$. The positron deposits energy and annihilates with an electron immediately, and the neutron is captured by Gd or H with a mean capture time of about 28 $\mu$s in the 0.1\% Gd-doped LS. The time and energy correlations are used to identify $\bar \nu_e$.

Similar correlated signals can be formed by $X\left(\alpha,n\right)Y$ reaction and mimic $\bar \nu_e$ events. In LS based experiment, the dominant reaction is ${^{13}}C\left(\alpha,n\right){^{16}}O$ since LS is mostly composed of $C_nH_m$ with very small fractions of Gadolinium, Nitrogen and Oxygen. The ${^{13}}C\left(\alpha,n\right){^{16}}O$ reaction forms the correlation as following: the prompt signal is neutron elastic scattering on proton or inelastic scattering on ${^{12}}C$, and the delayed signal is neutron capture on Gd or H. If ${^{16}}O$ is in the excited state, de-excitation of ${^{16}}O$ emits $\gamma$ or conversion electron which also contributes to the prompt signal. The signature is the same as $\bar \nu_e$, so such events can't be distinguished from $\bar \nu_e$ samples. The only solution is calculating the background rate and spectrum, then subtracting it statistically from $\bar \nu_e$ samples.

In pure LS, the dominant alpha source is ${^{210}}Po$, such as KamLAND~\cite{kamland}, while in Gd-LS, ${^{238}}U$, ${^{227}}Ac$ and ${^{232}}Th$ decay chains also contribute significantly, because U, Th and Ac can not be cleanly removed from Gd due to their similar chemical properties~\cite{LS}. The alpha rate in the detector can be determined by the Bi-Po cascade decays and the assumption of secular equilibrium of decay chains. The energy of alphas from natural radioactivity is on the order of several MeV, the energy distribution is shown in Fig.~\ref{alphaEnergy}.

\begin{center}
\includegraphics[width=7cm]{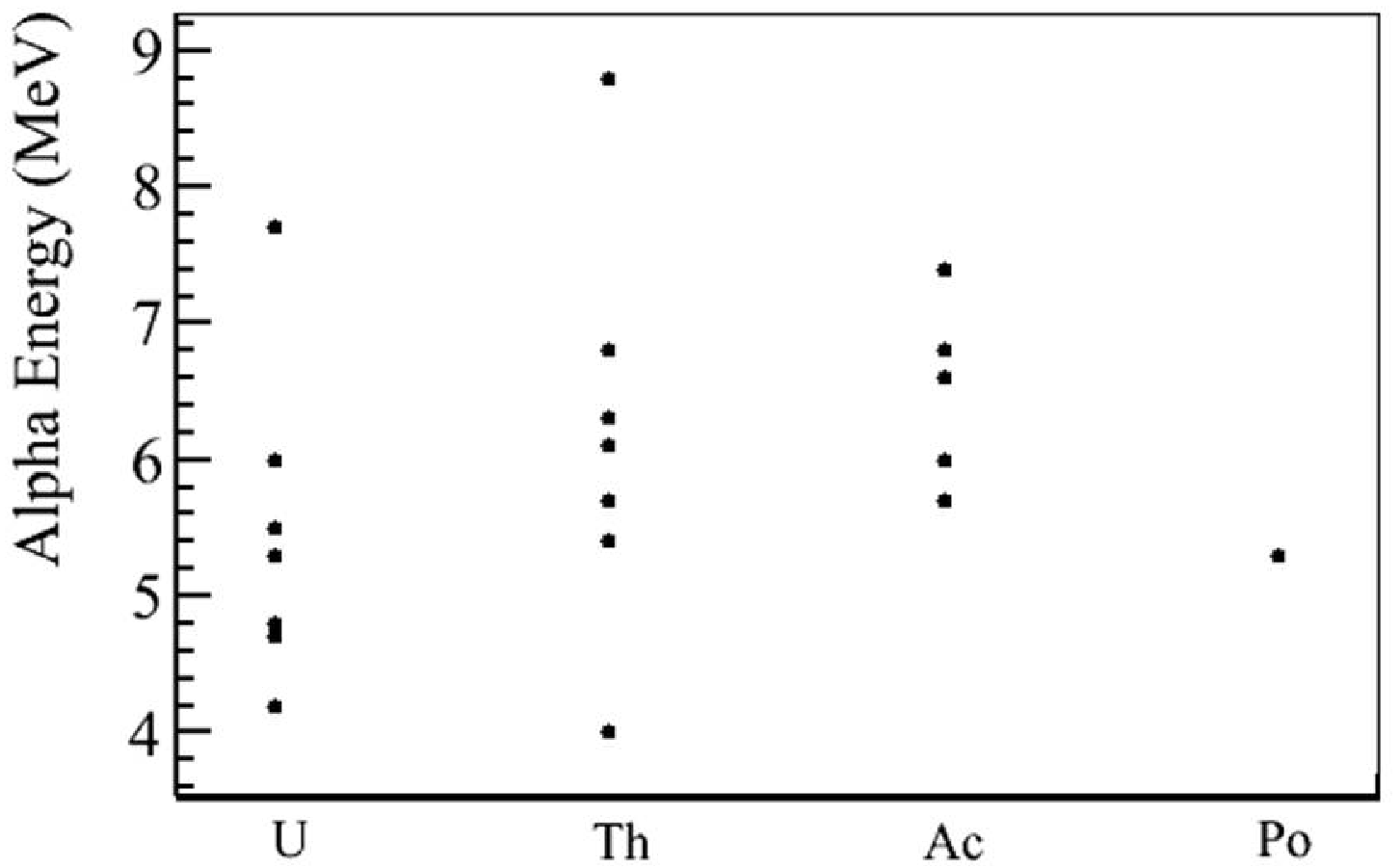}
\figcaption{\label{alphaEnergy}  Alpha energy distributions in the U, Th, Ac decay chains.  }
\end{center}

In this paper, the neutron yield and the calculation of the spectrum from the $^{13}C(\alpha,n)^{16}O$ reaction are described in detail. The estimated uncertainty is also discussed.

\section{Neutron yield calculation and uncertainty estimation}

\subsection{Neutron yield calculation}

 $\alpha$ interacts with ${^{13}}C$ and emits a neutron. The neutron yield is calculated from Eq.~(\ref{yield})

 \begin{eqnarray}
 N(E_{\alpha})&=&\int_{0}^{Range_\alpha}\sigma(E(x))Ndx\\
 &=&\int_{0}^{Range_\alpha}\sigma(E(x))\cdot\frac{\rho_{^{13}C}}{Z_{^{13}C}}\cdot N_A dx
 \label{yield}
 \end{eqnarray}

In the equation, where $\sigma(E(x))$ is the cross section depending on the $\alpha$ energy, $N$ is the number density of ${^{13}}C$, $Range_\alpha$ is the length that $\alpha$ can propagate in the detector. $N$ equals to $\frac{\rho_{^{13}C}}{Z_{^{13}C}}N_A$, where $Z_{^{13}C}$ is the atomic mass of ${^{13}}C$, $\rho_{^{13}C}$ is the density of ${^{13}}C$ in the detector, and $N_A$ is the Avogadro constant.

Because $\sigma(E(x))$, as shown in Fig.~\ref{CrossJE}, can not be analytically expressed as a function of $E$, the integration in Eq.~(\ref{yield}) should be done numerically, as shown in Eq.~(\ref{P(E)})

 \begin{equation}
  N(E_{\alpha})=\frac{\rho_{^{13}C}}{Z_{^{13}C}}\cdot N_A\cdot \sum_{step}\sigma(E_{step})d_{step}
  \label{P(E)}
 \end{equation}

where $d_{step}$ is the step length and sum of $d_{step}$ equals to $Range_\alpha$. $\sigma(E_{step})$ is the cross section at each step. To determine $d_{step}$ and $E_{step}$, GEANT4.9.2~\cite{GEANT} is utilized to simulate the $\alpha$ propagation in the detector.

The cross sections of the ${^{13}}C\left(\alpha,n\right){^{16}}O$ reaction is from JENDL~\cite{JENDL}\cite{JENDLBook} database, as shown in Fig.~\ref{CrossJE}. EXFOR database~\cite{exfor} is used for cross check, and will be discussed later.

As an example to estimate $\rho_{^{13}C}$, we assume the following properties using Daya Bay Gd-LS, which contains 87.7\% carbon content by weight, 12.1\% hydrogen, and 0.103\% Gd; the carbon density is 0.860$g/cm^3$. Since the natural abundance of ${^{13}}C$ is 1.1\%, $\rho_{^{13}C}$ is calculated as:

$\rho_{^{13}C} = 87.7\% \times 0.860 \times 1.1\% g/cm^3 = 0.0083g/cm^3$.

Knowing $d_{step}$, $E_{step}$, $\rho_{^{13}C}$ and $\sigma(E)$ in Eq.~(\ref{P(E)}), a Toy MC code is developed, in which the neutron yield of a certain $\alpha$ energy is calculated in the following steps:

(1) Start ToyMC, get the $d_{1}$, $E_{1}$ and $\sigma(E)$ of the first step;

(2) Calculate neutron yield in the step with Eq.~{\ref{P(E)}}, determine the fraction of ground and excited ${^{16}}O$ states;

(3) Subtract $E_{1}$ from $\alpha$ energy, get $d,E,\sigma$ of next step;

(4) Loop step 2 and 3 until $\alpha$ energy is zero;

(5) Sum the neutron yield of all steps.

The resulting neutron yields from the ${^{210}}Po$, ${^{238}}U$, ${^{232}}Th$ and ${^{227}}Ac$ decay chains are shown in Table~\ref{neutronYield}. The decay chains are assumed to be in equilibrium and the unit is neutron yield per decay of the chain. The uncertainties of the results will be discussed in the following section.

\begin{center}
\tabcaption{ \label{neutronYield}  Neutron yield per $^{210}Po$, $^{238}U$, $^{232}Th$ and $^{227}Ac$ decay. $N_{ground}$ refers to the neutron yield accompanied by a ground state $^{16}O$. $N_{excited}$ and $N_{total}$ have similar definitions.}
\footnotesize
\begin{tabular*}{80mm}{c@{\extracolsep{\fill}}ccccc}
\toprule DecayChain & $ N_{ground}$ & $N_{excited}$ & $N_{total}$ & Uncertainty \\
\hline
 ${^{210}}Po$ & 5.26e-8 & 4.90e-9 & 5.75e-8 & 7.2\%  \\
 ${^{238}}U$ & 4.34e-7 & 2.96e-7 &	7.30e-7 &	16.9\% \\
 ${^{232}}Th$ &	4.49e-7 & 4.92e-7 & 9.41e-7 & 27.7\% \\
 ${^{227}}Ac$ &	4.72e-7 & 6.18e-7 & 1.09e-6 & 25.9\% \\
\bottomrule
\end{tabular*}
\vspace{0mm}
\end{center}
\vspace{0mm}

The longer range associated with higher energies of $\alpha$s from the $^{227}Ac$ decay chain results in a larger neutron yield compared to $^{238}U$ and $^{232}Th$.

\begin{center}
\includegraphics[width=7cm]{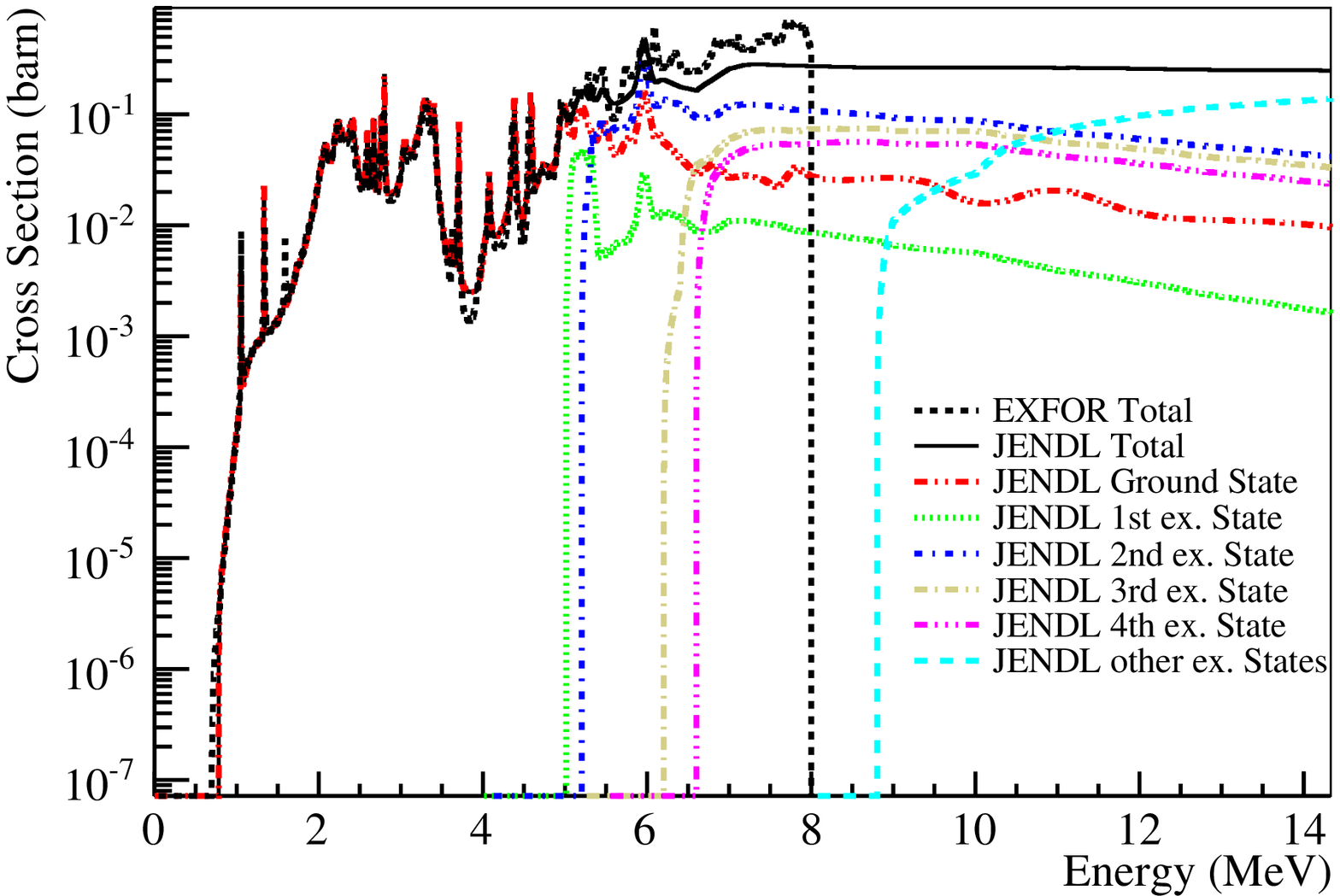}
\figcaption{\label{CrossJE}  Total cross section from JENDL and EXFOR. "ex." represents "excited". }
\end{center}

\subsection{Neutron yield uncertainty}

Because the values of $\rho_{^{13}C}$, $Z_{^{13}C}$, $N_A$ in Eq.~(\ref{P(E)}) are known precisely, uncertainties arise from: (1) cross section; (2) numerical integration.

In the above calculation the JENDL database is used. To estimate the uncertainty associated with the cross sections, the results were cross-checked with inputs from the EXFOR database, as shown in Fig.~\ref{CrossJE}. The EXFOR database provided the absolute cross section of the $^{13}C(\alpha,n)^{16}O$ reaction at $E_{\alpha}$=0.8 to 8.0 MeV. Below 5.6 MeV, both data sets are in good agreement while the cross sections exhibit some tensions at larger $\alpha$ energies.

Since both of the two databases claim their cross sections were determined based on experimental data and are in good agreement with experimental data~\cite{JENDLBook}\cite{exfor}, it is difficult to make a conclusion which database is right. So we utilize the two databases to calculate two neutron yields, and the difference is treated as the systematic uncertainty. The results are shown in Table~\ref{yieldError1}, and the definition of systematics is $|N_{JENDL}-N_{EXFOR}|/N_{JENDL}$.

\begin{center}
\tabcaption{ \label{yieldError1} Neutron yield differences with two cross section databases.}
\footnotesize
\begin{tabular*}{80mm}{c@{\extracolsep{\fill}}cccc}
\toprule DecayChain & $N_{JENDL}$ & $N_{EXFOR}$ & Systematics \\
\hline
 ${^{210}}Po$ & 5.75e-8 & 5.37e-8 & 6.6\%  \\
 ${^{238}}U$ & 7.30e-7 & 8.51e-7 & 16.6\% \\
 ${^{232}}Th$ & 9.41e-7 & 1.20e-6 & 27.5\% \\
 ${^{227}}Ac$ & 1.09e-6 & 1.37e-6 & 25.7\% \\
\bottomrule
\end{tabular*}
\vspace{0mm}
\end{center}
\vspace{0mm}

In the numerical integration, GEANT4 is used to calculate the $\alpha$ range, $d_{step}$ and dE/dx. To determine the associated uncertainty, $d_{step}$ is varied and SRIM~\cite{SRIM} is employed to validate the $\alpha$ range and dE/dx.

SRIM is a collection of software packages which calculates many features of ion transport in matter~\cite{SRIM}. It can provide $\alpha$ range and dE/dx, and comparison between GEANT4 and SRIM is shown in Fig.~\ref{dEdX}. Differences are small in the high energy region.

\begin{center}
\includegraphics[width=7cm]{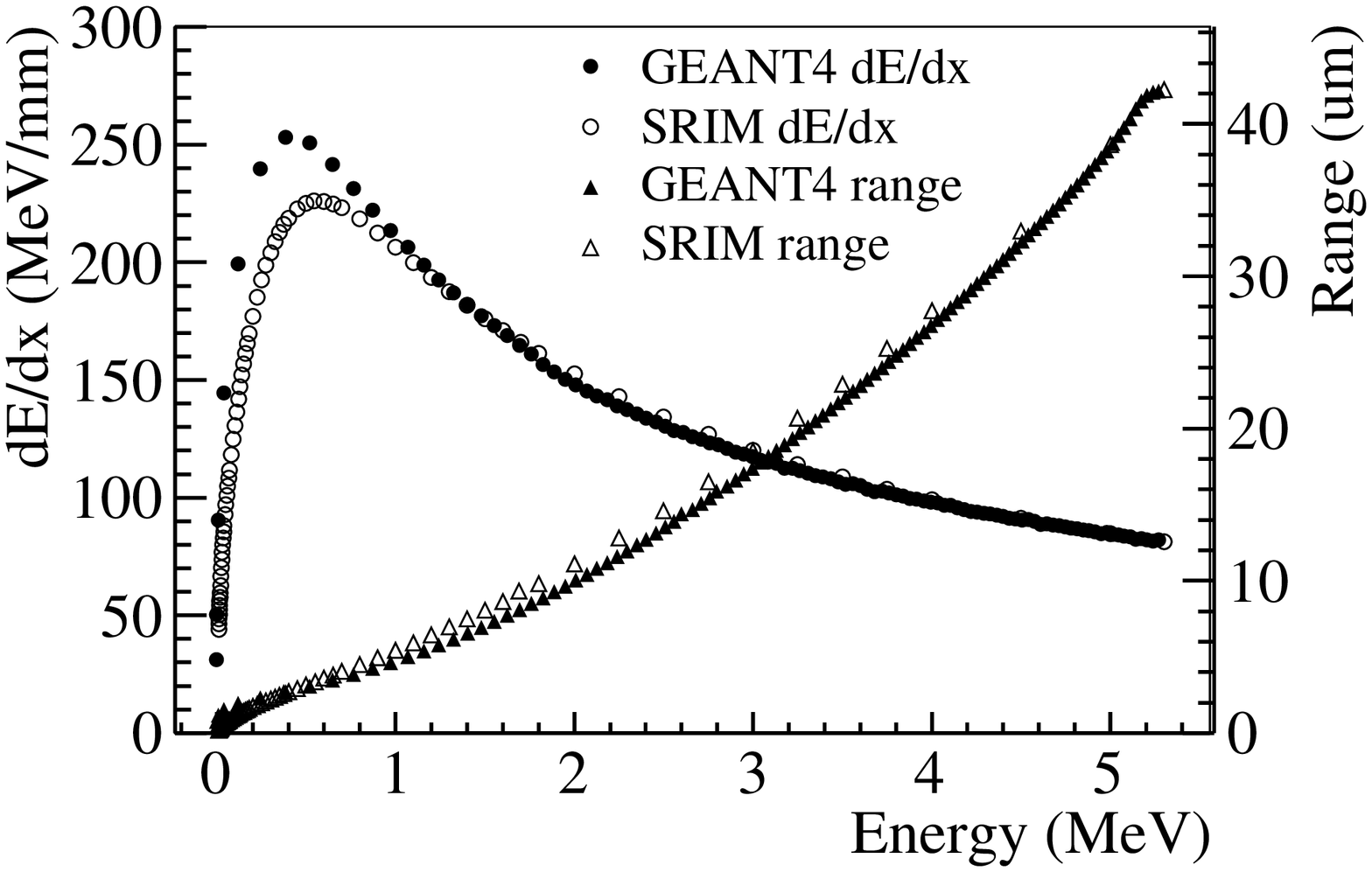}
\figcaption{\label{dEdX}  The dE/dx and $\alpha$ range comparison between GEANT4 and SRIM.}
\end{center}

With a given 5.3 MeV $\alpha$ range and dE/dx, we divide the range uniformly into 200 to 5000 steps, and compare the neutron yield with the GEANT4 step division result (106 steps, the first point in the plot), as shown in Fig.~\ref{StepGS}. The yield uncertainty is discussed as following:

(1) We adopt a piecewise formula of Newton-Cotes quadrature as the numerical integration method. The truncation error is on the order $O(h^2)$(h is the step length), when h is small enough, the numerical result converges to the real value of the integration. The middle panel of Fig.~\ref{StepGS} shows the relative difference of GEANT4 step lengths and uniform divisions. Differences from 3000 steps to 5000 steps are less than 0.1\%, which can be tolerated, so we consider the result of 5000 steps as accurate. Since we calculate neutron yield with GEANT4 step lengths, the difference of GEANT4 result and 5000 steps result is considered as the systematic uncertainty: 2.5\%.

(2) As shown in the bottom panel of Fig.~\ref{StepGS}, the difference in neutron yield calculated using GEANT4 and SRIM is about 1.5\%, independent of the number of steps. So 1.5\% is taken as the systematic uncertainty.

\begin{center}
\includegraphics[width=7cm]{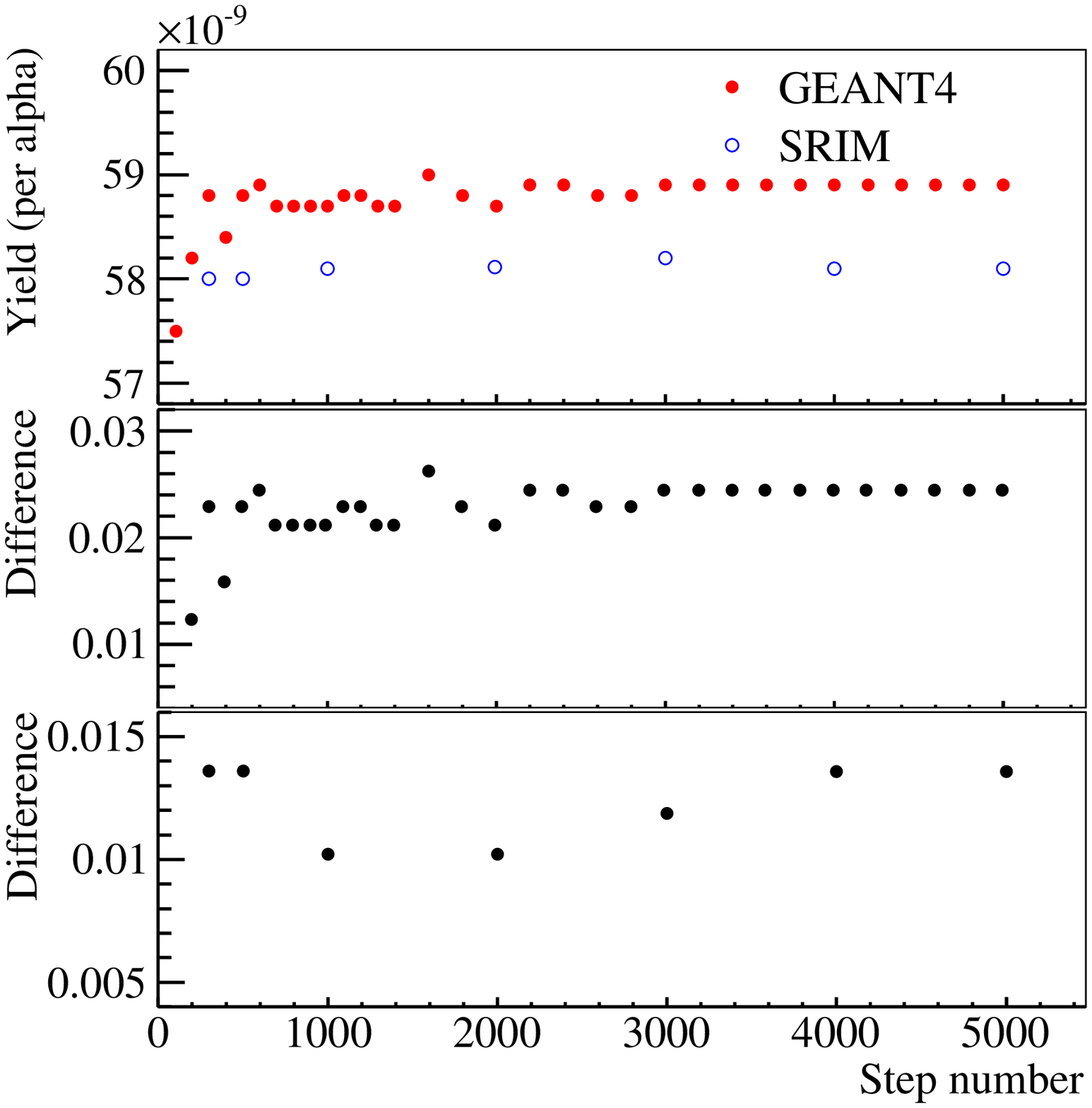}
\figcaption{\label{StepGS}  Top Panel: neutron yield with different step lengths and different software, the first point of GEANT4 uses step lengths given by GEANT4 itself, and other points divide the range uniformly to certain steps; Middle Panel: relative difference of GEANT4 step lengths and uniform divisions; Bottom Panel: relative difference of GEANT4 and SRIM. }
\end{center}

In conclusion, the transportation model introduces 2.9\% systematic uncertainty. Combining with the cross section uncertainties in Table~\ref{yieldError1}, the total uncertainties are estimated in the last column of Table~\ref{neutronYield}. Uncertainty of $^{238}U$, $^{232}Th$ and $^{227}Ac$ decay chains are larger than $^{210}Po$, because the former three chains contains high energy $\alpha$, and at high energy region, JENDL and EXFOR have large discrepancy. When applying this calculation to a specific experiment, the uncertainty of the $\alpha$ rate estimation should also be taken into consideration.

The calculation is compared with KamLAND results~\cite{kamland}, in which the dominant $\alpha$ source is $^{210}Po$. The KamLAND LS components and alpha rate are known, then the alpha-n background is calculated using the above method with a result of 16.9, consistent with the KamLAND result 17.8$\pm$7.3.

\section{Background spectrum and uncertainty}

The background spectrum consists of: (1) neutron elastic recoil on proton; (2) de-excitation of ${^{16}}O$; (3) $\alpha$ deposited energy before the reaction. In this section, the spectrum and uncertainty will be discussed.

\subsection{Spectrum determination}

The neutron energy depends on the $\alpha$ energy before the reaction and the final state of ${^{16}}O$. When a neutron is generated, the $\alpha$ deposited energy is recorded. According to the energy, momentum conservation law, the neutron kinetic energy in the center-of-mass (CM) frame is calculated and then converted to the laboratory (Lab) frame:

\begin{eqnarray}\label{CM}
E_{n,lab} &=& E_{n,cm}+{\frac{{m_1}\times{m_3}}{{(m_1+m_2)}^2}}\times{E_{a,lab}}+\nonumber\\
&&{\frac{2\sqrt{(m_1m_3E_{a,lab}E_{n,cm})}}{m_1+m_2}}\times{cos(\theta_c)}.
\end{eqnarray}

where $E_{n,cm}$ is the CM neutron energy, $m_1$ is the mass of $\alpha$, $m_2$ is mass of the ${^{13}}C$, $m_3$ is the mass of neutron and $E_{\alpha,lab}$ is $\alpha$ energy in the Lab frame. $\theta_c$ is neutron scattering angle in the CM frame, which is randomly sampled in [0,$\pi$].

\begin{center}
\includegraphics[width=7cm]{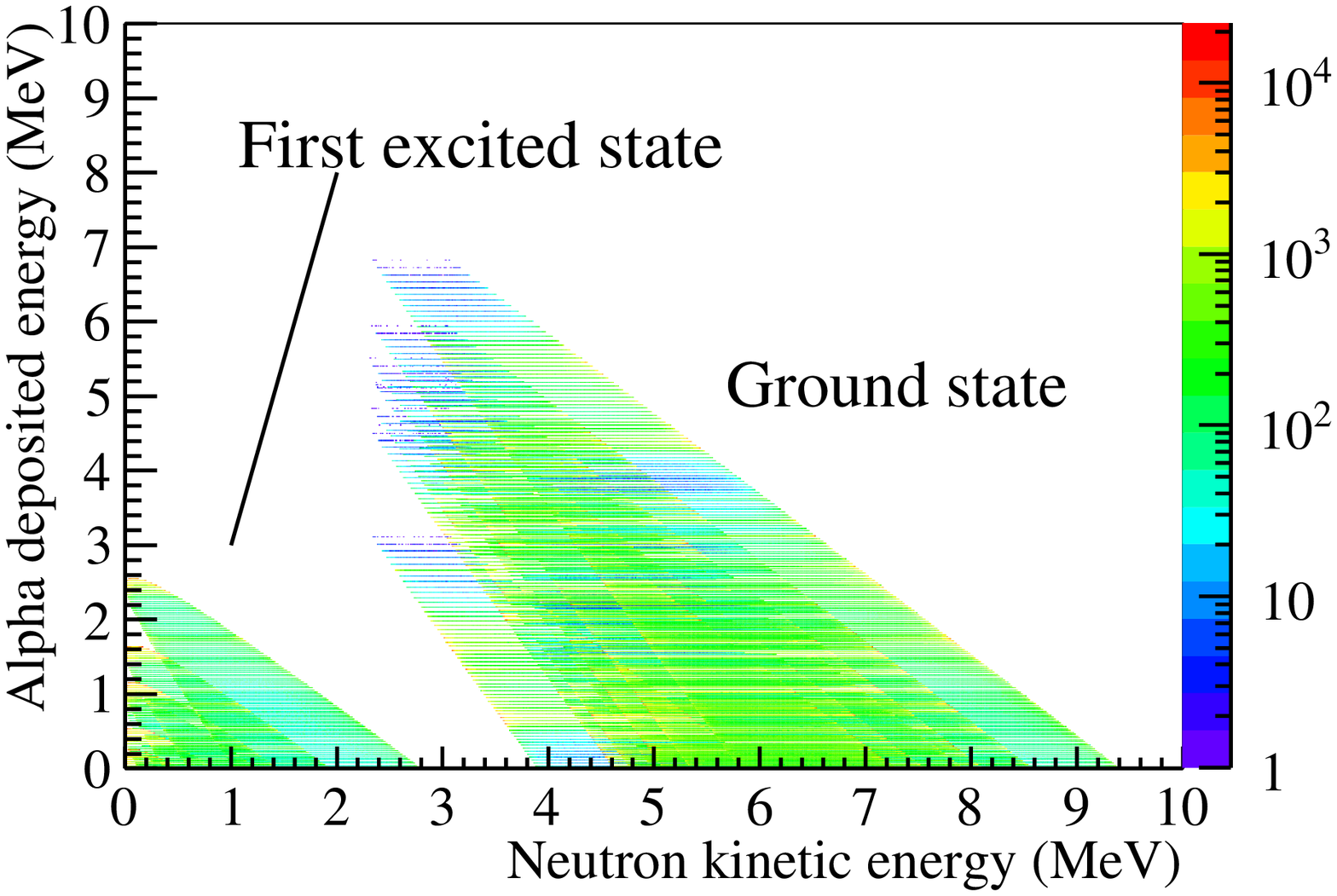}
\figcaption{\label{two dimension}  $\alpha$ deposited energy versus neutron kinetic energy distribution from $^{232}$Th.   }
\end{center}

Fig.~\ref{two dimension} shows the 2-dimensional probability density function (PDF) in $\alpha$ energy and neutron kinetic energy. The kinematics and step by step transportation simulation was used in generating this figure, and the color coding in Z encodes the reaction probability. A larger initial $\alpha$ energy into the reaction results in a larger neutron kinetic energy distribution. Neutrons accompanied by $^{16}O$ excited states have much lower kinetic energies than those by $^{16}O$ ground state, because $^{16}O$ excited states require more than 6 MeV energy to be excited.

 In order to determine the background spectrum for a given experiment, the 2-dimensional histogram can be used as input for a MC simulation. The Daya Bay detector simulation is used to illustrate the procedure outlined in the following steps:

(1) For a given $^{16}O$ final state, sample alpha and neutron energy using the PDF in Fig.~\ref{two dimension}. In Gd-LS based detectors, protons from neutron scattering and alphas are quenched, with typical quenching factors of around 0.2 and 0.1 for 5 MeV protons and alphas, respectively. The visible energy is therefore smaller than the kinetic energy.

(2) Add the energies from the de-excitation of the $^{16}O$ excited states. These include the 6.13 MeV, 6.92 MeV, 7.11 MeV $\gamma$ for the $2^{nd}$, $3^{rd}$, $4^{th}$ excited states, respectively, and a 6.03 MeV $e^+e^-$ pair from the $1^{st}$ state.

(3) Generate the vertex combining the particles in (1) and (2). Alpha slowing down, neutron scattering and proton slowing down, $^{16}O$ decay and gamma energy deposit happen at nanosecond time scale, and all of them contribute to the prompt energy spectrum.

(4) Combine the results for different channels, using the overall integrated probabilities.

The resulting total background spectra of the four decay chains are shown in Fig.~\ref{background spectrum}. The calculated spectrum changes based on the mean $\alpha$  energy from the respective decay chain, with the ratio of the $^{16}O$ excited state and the associated peak around 6.5 MeV increasing with larger $\alpha$ energies. The peak around 5 MeV is introduced by neutron inelastic scattering of $^{12}C$.

\begin{center}
\includegraphics[width=7cm]{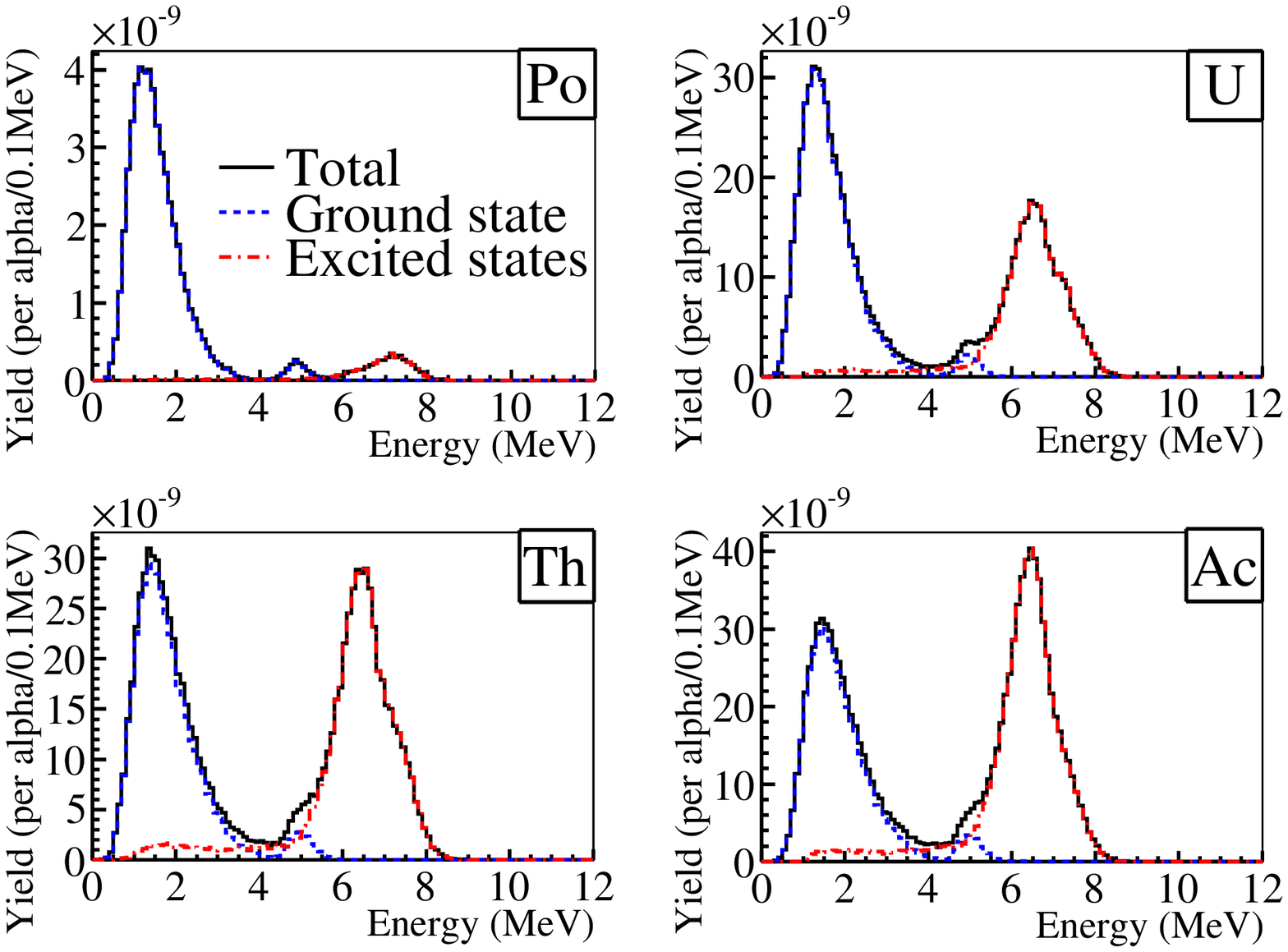}
\figcaption{\label{background spectrum}  Visible energy spectra of backgrounds from $^{210}Po$, $^{238}U$, $^{232}Th$ and $^{227}Ac$. }
\end{center}

\subsection{Spectrum uncertainty}

Similar to the neutron yield uncertainty, spectrum uncertainties arise from: dE/dx and $\alpha$ range, the $\left(\alpha,n\right)$ angular distribution and cross section.

SRIM is used again to validate the GEANT4 results. Neutron kinetic energy spectrum comparison between GEANT4 and SRIM of 5.3 MeV $\alpha$ is shown in Fig.~\ref{KineticJS}, and the difference is negligible.

\begin{center}
\includegraphics[width=7cm]{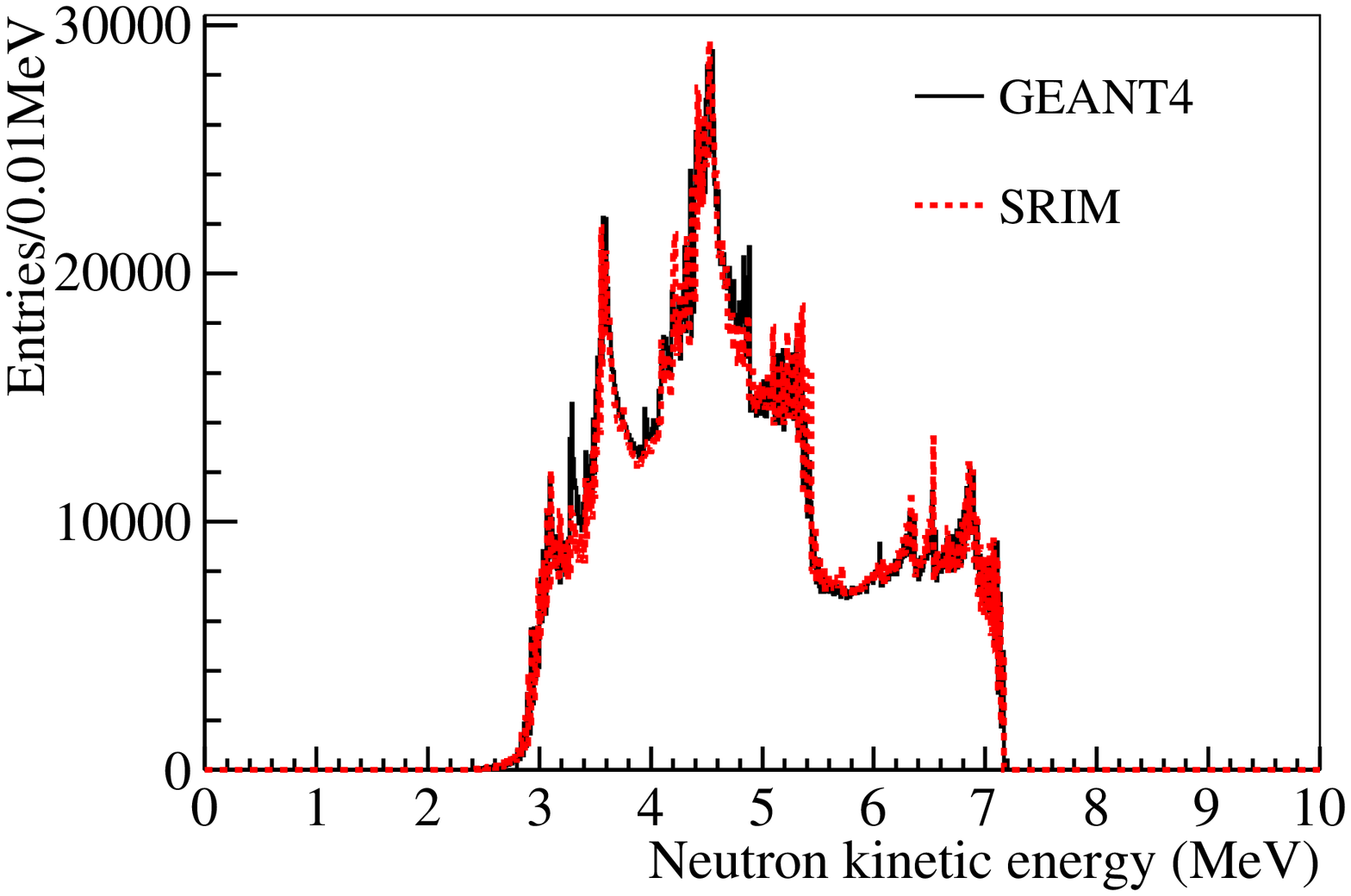}
\figcaption{\label{KineticJS}  Neutron kinetic energy spectrum comparison between GEANT4 and SRIM. }
\end{center}

 In Eq.~\ref{CM}, neutron scattering angle could not be precisely calculated, and common methods are uniformly sampling $\theta_c$ in [0,$\pi$] (assuming a $\frac{d\sigma}{d\Omega}\thicksim\frac{1}{sin\theta_c}$) or $cos\theta_c$ in [-1,1] (isotropic angular distribution in $\frac{d\sigma}{d\Omega}$). Difference of neutron kinetic spectrum calculated using the two sampling methods is very small. In order to crosscheck the JENDL result, the EXFOR database is used. Neutron kinetic spectrum obtained using the two databases are almost the same.

Finally, the above neutron kinetic spectra with same statistics are sent to MC, and visible energy spectra are shown in the top panel of Fig.~\ref{spectrumError}. The "JENDL+theta" is the default configuration which we used in the last section, and "JENDL+isotropic" is the spectrum with different $\theta$ sampling method, and "EXFOR+theta" is the spectrum with different database. Spectra are compared as shown in the middle panel and the bottom panel. Differences are related to energy, and it is difficult to give a unified uncertainty. So an error band is drawn on the spectrum of the top panel, and error of each bin is estimated by the sqrt quadratic sum of the differences in the middle and bottom panel.

\begin{center}
\includegraphics[width=7cm]{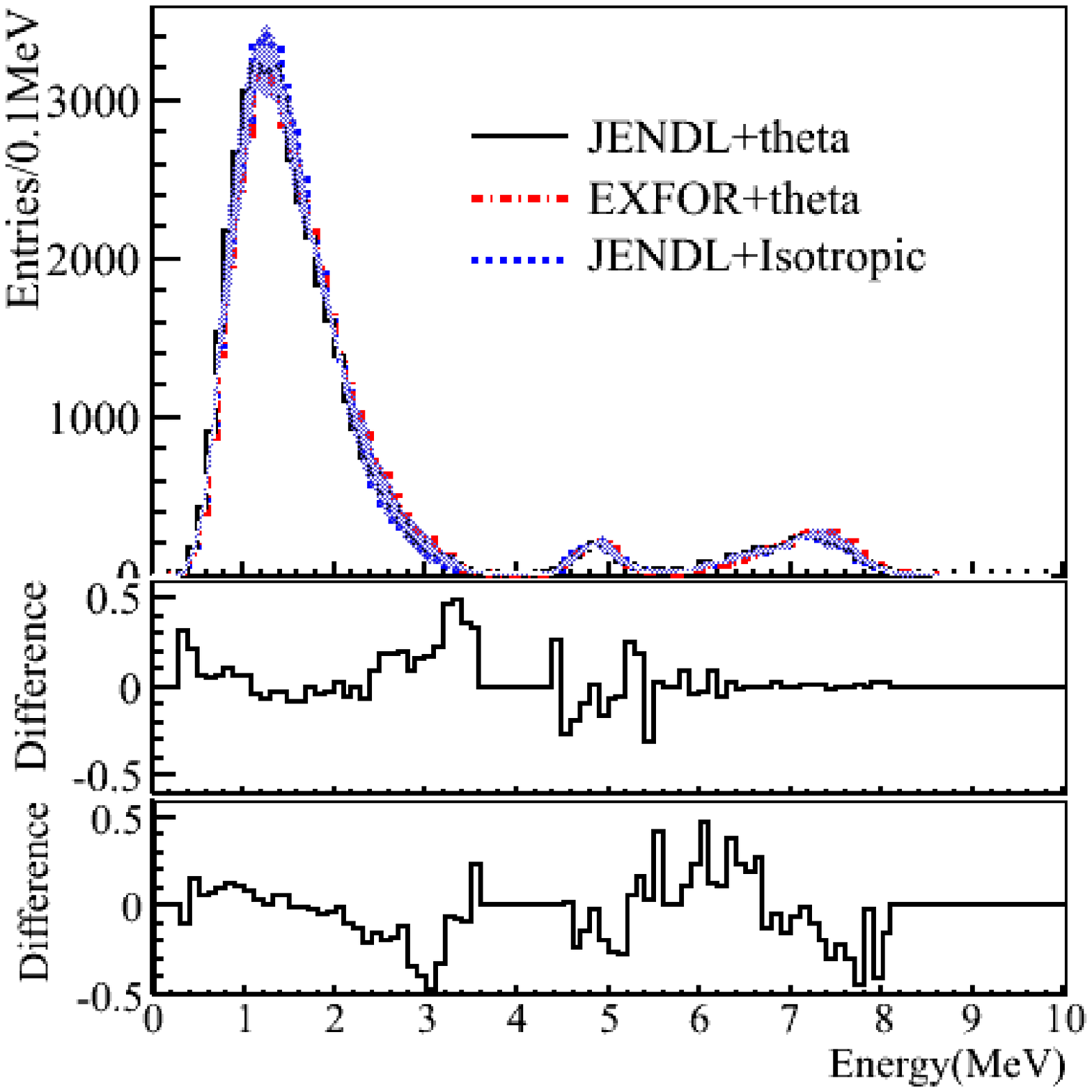}
\figcaption{\label{spectrumError}  The top panel shows the neutron visible energy spectra derived using different methods. The relative difference of the spectra based on different $\theta$ sampling is depicted in the middle panel. The bottom panel is comparison of different databases. Combining the comparison, an error band is given in the top panel.}
\end{center}

To a certain experiment, the discrepancy between data and MC for neutron simulation should also be considered as a spectrum uncertainty source. Since results in the paper assume no geometric effects, for $^{210}Po$ attached on the acrylic vessel, one can not directly use the results in Table~\ref{neutronYield} and Fig.~\ref{background spectrum}.


\section{Conclusion}

The ${^{13}}C\left(\alpha,n\right){^{16}}O$ neutron yield in liquid scintillator detector is calculated for internal U/Th/Ac/Po backgrounds with estimated uncertainty. We also estimate the prompt visible energy spectrum with uncertainties. The results can be directly applied to estimate background in the current and future Gd-LS or LS based reactor neutrino experiments.

\section{Acknowledgement}
\acknowledgments{The experiment is supported by the Ministry of Science and Technology of the People's Republic of China (2013CB834301). The authors would like to acknowledge the Daya Bay collaborators, especially Dr. Miao He, Dr. Chao Zhang, Dr. Soeren Jetter and Dr. Liang Zhan, for helpful comments.}




\end{multicols}

\clearpage

\end{CJK*}
\end{document}